\documentclass[]{aa}
\usepackage{natbib}
\usepackage{epsfig}
\def\ut#1{\mathop{\vtop{\ialign{##\crcr
     $\hfil\displaystyle{#1}\hfil$\crcr\noalign
     {\kern1pt\nointerlineskip}\hbox{$\hfil\sim\hfil$}\crcr
     \noalign{\kern1pt}}}}}

\def\undersymbol#1#2{\mathop{\vtop{\ialign{##\crcr
     $\hfil\displaystyle{#2}\hfil$\crcr\noalign
     {\kern1pt\nointerlineskip}\hbox{$\hfil#1\hfil$}\crcr
     \noalign{\kern1pt}}}}}

\begin{document}

\title{A new analysis of the MEGA M31 microlensing events}

\author{G. Ingrosso\inst{1},
        S. Calchi Novati\inst{2},
        F. De Paolis\inst{1},
        Ph. Jetzer\inst{3},
        A.A. Nucita\inst{1},
        G. Scarpetta\inst{2},
        F. Strafella\inst{1}}
\institute{Dipartimento di Fisica,
           Universit\`{a} di Lecce and INFN, Sezione di Lecce,
           CP 193, I-73100 Lecce, Italy \and
           Dipartimento di Fisica ''E. R. Caianiello'',
           Universit\`{a} di Salerno, I-84081 Baronissi (SA)
           and INFN, Sezione di Napoli, Italy \and
           Institute for Theoretical Physics,
           University of  Z\"{u}rich, Winterthurerstrasse 190,
           CH-8057 Z\"{u}rich, Switzerland}

\date{Received  / Accepted}
\authorrunning{Ingrosso et al.}
\titlerunning{Microlensing events towards M31}
\abstract{
We discuss the results of the MEGA microlensing campaign
towards M31. Our analysis is based on an analytical evaluation
of the microlensing rate, taking into account
the observational efficiency as given by the MEGA collaboration.
In particular, we study the spatial and time duration
distributions of the microlensing events for several
mass distribution models of the M31 bulge.
We find that only for extreme models of the M31 luminous
components it is possible to reconcile the total observed MEGA events
with the expected self-lensing contribution.
Nevertheless, the expected
spatial distribution of self-lensing events
is more concentrated and hardly in agreement with the observed distribution.
We find it thus difficult to explain all events as being due
to self-lensing alone. On the other hand, the small number of events
does not yet allow to draw firm conclusions on the halo dark matter
fraction in form of MACHOs.

}

\maketitle

\section{Introduction}
Since the proposal of \cite{pacz86} gravitational
microlensing has probed to be an efficient
tool for the study of the MACHO contribution
to the dark matter galactic halos.
The first line of sight to be explored with this purpose
has been that towards the Magellanic Clouds \citep{macho93,eros93,ogle93}.
As first discussed by \cite{crotts92}, \cite{agape93} and \cite{jetzer94}
observations towards M31 have also been undertaken \citep{crotts96,agape97}.

The interpretation of the results obtained so far remain, however,
debated and controversial. Along the line of sight towards the LMC
the MACHO collaboration \citep{MACHO00} reported the signal of a
halo fraction of about 20\% in form of MACHOs with mass $\simeq
0.5~ M_{\odot}$, while the latest results of the EROS
collaboration towards both the SMC and the LMC
\citep{eros03,eros05,eros06} are even compatible with a no MACHO
hypothesis.

The case towards M31 is complicated by the further degeneration in
the lensing parameter space due to the fact that sources at
baseline are unresolved, a case referred to as ``pixel-lensing''
\citep{crotts92,agape93,gould96}. Still, a handful of microlensing
events have been observed in the meantime
\citep{agape99,auriere01,novati02,novati03,paulin03,wecapp03,
mega04,belokurov05,joshi05} and lately the first constraints on
the halo fraction have been reported. The results of the
POINT-AGAPE collaboration \citep{novati05} are compatible with the
ones of the MACHO group, by putting a lower limit on the halo
fraction in form of MACHOs of $\sim 20\%$ for objects in the mass
range $0.5 - 1~ M_{\odot}$. On the contrary the MEGA collaboration
\citep{mega05} finds that their results, although not conclusive,
are in agreement with a no MACHO hypothesis. Although the issues
involved in the microlensing observations towards the LMC or the
M31 are indeed rather different, the results on the halo fraction
in form of MACHOs depend crucially on the prediction of the
expected signal due to known luminous populations, this being
dominated by the ``self-lensing'' signal where both source and
lens belong to same star population residing respectively either
in the LMC or in M31. This problem is indeed the main aspect we
want to discuss in this paper.

The issue of the expected microlensing signal
towards M31 has been discussed in a few works
(e.g. \citealt{kerins01,baltz03,wecapp05,kerins06}).
In this respect the modeling of the M31 luminous components is a crucial
aspect to be dealt with in order to get meaningful results.
We have first considered these aspects in \cite{depaolis05},
then, more recently, in \cite{ingrosso06} we have
developed a Monte Carlo simulation which
we have used to investigate the nature and location
of the microlensing candidates events towards M31 as reported
in a first paper by the MEGA collaboration \citep{mega04}.
In the present work our aim is to further
explore these issues taking into account the latest MEGA results \citep{mega05}.
In particular, we now go through a full characterisation
of the expected signal, including the predicted number of events,
that we then compare with the observational results. Our aim
is to explore, in particular, the question whether
the expected self-lensing signal due to stars
belonging either to the bulge or the disc of M31 is able,
as claimed by \cite{mega05}, to fully explain their results.

The plan of the paper is as follows. In Sect. 2
we describe the microlensing rate,
our main tool of investigation, and present its predictions.
In Sect. 3 we critically discuss the models used to describe
the different galactic components involved.
In Sect. 4 we discuss our main results and give some concluding
remarks.

\section{Event rate prediction}

In evaluating \footnote{Here we follow with some modifications the
derivation in our previous paper \cite{ingrosso06}.} the expected
event number along a fixed line of sight we take into account the
existence of two source populations (stars in the M31 bulge and
disk) with number density $n_{\mathrm s}(D_{os}, M)$ and of five
lens populations (stars in the M31 bulge, stars in the M31 and MW
disks, MACHOs in M31 and MW halos) with density $n_{\mathrm
l}(D_{ol},\mu)$. Here $D_{os}$ ($D_{ol}$) is the source (lens)
distance from the observer, $M$ is the source magnitude and $\mu$
is the lens mass in solar units.

We assume, as usual, that the mass distribution of the lenses
is independent on their position in M31 or in the Galaxy
({\it factorization hypothesis}). So, the lens number density
(per unit of volume and mass) $n_{\mathrm l} (D_{ol}, \mu)$ can be
written as \citep{jms}
\begin{equation}
n_{\mathrm l}(D_{ol}, \mu)=
\left( \frac{ \rho_{\mathrm l}(D_{ol})} {\rho_{\mathrm l} (0)} \right)
 ~ \psi_0(\mu)~,
\label{nnll}
\end{equation}
where  $\rho_{\mathrm l}(0)$ is the local mass density of the considered
lens population in the Galaxy
or the central density in M31,
$\psi_0(\mu)$ the corresponding
lens number density per unit mass and
the normalization is given by
\begin{equation}
\int_{\mu_{\mathrm{min}}}^{\mu_{\mathrm{up}}}
\psi_0(\mu) ~ \mu ~d\mu~ = \frac{\rho_{\mathrm l}(0)}{M_{\odot}}~.
\label{nfm}
\end{equation}
Here $\mu_{\mathrm {min}}$ and $\mu_{\mathrm {up}}$ are the lower and the
upper limits for the lens masses (see Subsection 3.2).

Likewise, assuming that the magnitude distribution of the sources
is independent on their position in M31,
the source number density (per unit of volume and magnitude)
$n_{\mathrm s} (D_{os},M)$ can be written as
\begin{equation}
n_{\mathrm s}(D_{os}, M)=
\left( \frac{ {\cal{L}}_{\mathrm s}(D_{os}) }
            { {\cal{L}}_{\mathrm s}(0)      } \right)
~\phi_{\mathrm s} (M)~,
\label{phi0}
\end{equation}
where ${\cal{L}}_{\mathrm s}(0)$ is central luminosity density
of the considered source population,
$\phi_{\mathrm s} (M)$
is the source number density per unit magnitude
in the M31 center and the normalization now reads
\begin{equation}
\int_{M_{\mathrm {min}}}^{M_{\mathrm{up}}}
\phi_{\mathrm s} (M)~L(M)~ dM~ = {\cal{L}}_{\mathrm s}(0)~.
\label{normphi}
\end{equation}
\begin{table}
\caption{For the 14 MEGA events we give position,
magnitude at maximum $\Delta r$ and full-width half-maximum
duration $t_{1/2}$.
The coordinate system we adopt
has origin in the M31 center and the
X axis oriented along the M31 disk major
axis.}
\medskip
\begin{tabular}{|c|c|c|c|c|}
\hline
MEGA & X        & Y        & $\Delta r$ & $t_{1/2}$  \\
     & arcmin   &  arcmin  & mag        & day       \\
\hline
    1      & -4.367 &  -2.814& $21.8  \pm 0.4 $ & $ 5.4 \pm 0.7$ \\
    2      & -4.478 &  -3.065& $21.51 \pm 0.06$ & $ 4.2 \pm 0.7$\\
    3      & -7.379 &  -1.659& $21.6  \pm 0.1 $ & $ 2.3 \pm 2.9$\\
    7 (N2) & -21.164&  -6.248& $19.37 \pm 0.02$ & $17.8 \pm 0.4$\\
    8      & -21.650&  +7.670& $22.3  \pm 0.2 $ & $27.5 \pm 1.2$\\
    9      & -33.833&  -2.251& $21.97 \pm 0.08$ & $ 2.3 \pm 0.4$\\
   10      &  -3.932& -13.847& $22.2  \pm 0.1 $ & $44.7 \pm 5.6$\\
   11 (S4) & +19.193& -11.833& $20.72 \pm 0.03$ & $2.3  \pm 0.3$\\
   13      & +22.072& -22.022& $23.3  \pm 0.1 $ & $26.8 \pm 1.5$\\
   14      & +19.349& -29.560& $22.5  \pm 0.1 $ & $25.4 \pm 0.4$\\
   15      &  -6.634& -0.697 & $21.63 \pm 0.08$ & $16.1 \pm 1.1$\\
   16 (N1) &  -6.886& +3.843 & $21.16 \pm 0.06$ & $1.4  \pm 0.1$\\
   17      & +21.214&  -5.161& $22.2  \pm 0.1 $ & $10.1 \pm 2.6$\\
   18      &  +6.995& -13.533& $22.7  \pm 0.1 $ & $33.4 \pm 2.3$\\
\hline
\end{tabular}
\label{tab0}
\end{table}
Here $M_{\mathrm {min}}$ and $M_{\mathrm {up}}$ are the lower and the
upper limits for the source magnitude (see Subsection 3.1),
$L(M)$ is the luminosity in a given photometric band
\begin{equation}
L(M) = \eta_{Vega}~L_{\odot} ~ 10^{-M/2.5}~,
\end{equation}
$\eta_{Vega}$ being the Vega luminosity (in solar units) in the considered
band.

We consider the volume element of the microlensing tube to be $d^3
x = (\vec v_{\mathrm{r \perp}} \cdot \vec n) R_E u_{\mathrm th}
d\alpha dD_{\mathrm{ol}}$, $R_E$ being the Einstein radius, $\vec
v_{\mathrm r \perp}$ the relative tranverse velocity between the
lens and the microlensing tube with distribution function
$f(\vec{v}_{\mathrm{r \perp}})$, $\vec n$ the inner normal to the
microlensing tube and $\alpha$ the angle between $\vec n$ and
$\vec {A}_{\perp}$ (see eq. \ref{aaa}). Assuming perfect
observational sensitivity to microlensing, the differential event
rate $dN_{\mathrm{ev}}/d\Omega$ (in units of event sr$^{-1}$) for
microlensing by compact objects with impact parameter below a
certain threshold $u_{\mathrm th}$, during the time interval $dt$,
is given by \citep{Griest,djm}
\begin{eqnarray}
\label{dnev}
\frac{dN_{\mathrm{ev}}}{d \Omega} & = &
D^2_{os}~ u_{\mathrm th} R_E d\alpha ~
v_{\mathrm r \perp} f(\vec v_{\mathrm{r \perp}}) d^2 \vec v_{\mathrm r \perp}
\cos \theta ~ \\ \nonumber
& &
n_{\mathrm l}(D_{ol},\mu) ~ n_{\mathrm s}(D_{os},M)
~ d\mu ~ dM ~ dD_{\mathrm{os}}~  dD_{\mathrm{ol}}~dt~,
\end{eqnarray}
where $\theta \in (-\pi/2,\pi/2)$ is the angle between
$\vec n$ and $\vec v_{\mathrm r \perp}$.
We assume that the velocity distributions of lenses and sources
are isotropic around their streaming velocities (if present)
due to the rotation of the considered population with respect to the
M31 or MW center
(we neglect any transverse
drift velocity of the M31 center with respect to the Galaxy).
Accordingly, the lens (source) velocity is splitted into
a random component - which follows a Maxwellian distribution with
one-dimensional velocity dispersion
$\sigma_{\mathrm{l}}$ ($\sigma_{\mathrm{s}}$) -
and a streaming component, namely
$\vec{v}_{\mathrm{l}} = \vec{v}_{\mathrm{l,ran}}+\vec{v}_{\mathrm{l,rot}}$
and
$\vec{v}_{\mathrm{s}} = \vec{v}_{\mathrm{s,ran}}+\vec{v}_{\mathrm{s,rot}}$.
When the lens and source velocities are projected in the lens plane
(transverse to the microlensing tube), the respective random velocity
distributions are again described by Maxwellian functions,
with the same one-dimensional velocity dispersion
$\sigma_{\mathrm{l}}$ for lenses, and with (projected) dispersion
$(D_{\mathrm l}/D_{\mathrm s}) \sigma_{ \mathrm s}$ for sources.
Then, neglecting the streaming,
the relative, projected, random  velocity
$\vec{v}_{\mathrm{ls \perp,ran}}=
 \vec{v}_{\mathrm{l \perp,ran}}-
 \vec{v}_{\mathrm{s \perp,ran}}$
of lenses and sources
is a maxwellian distribution
$f(\vec{v}_{\mathrm{ls \perp,ran}})$ with combined width
\begin{equation}
\sigma_{\mathrm {sl}}  = \sqrt{ \sigma_{\mathrm l}^2 +
\left(D_{\mathrm {ol}}/D_{\mathrm {os}}\right)^2
\sigma_{\mathrm s}^2}~.
\label{sigmacomposta}
\end{equation}

We now include all streaming motions in the
vector $\vec A_{\perp}$ defined as the difference between the
projected, streaming velocities of lenses, sources and observer,
namely
\begin{equation}
\vec {A}_{\perp} =
\left(1- \frac{D_{\mathrm{ol}}}{D_{\mathrm{os}}}\right)
\vec{v}_{\mathrm \sun \perp,rot}+
\left(   \frac{D_{\mathrm{ol}}}{D_{\mathrm{os}}}\right)
\vec{v}_{\mathrm {s}  \perp,rot}
 -                                   \vec{v}_{\mathrm {l}  \perp,rot}~.
\label{aaa}
\end{equation}
The resulting distribution function $f(\vec{v}_{\mathrm{r \perp}})$
of the relative, transverse
velocity between the lenses and the microlensing
tube is now given by the Maxwellian function
$f(\vec{v}_{\mathrm{ls \perp,ran}})$ shifted by the vector
$\vec{A}_{\perp}$, that we write in polar coordinates on the lens plane
as
\begin{equation}
\label{vr_distribution}
\displaystyle{
f(\vec{v}_{\mathrm{r \perp}}) d^{2} \vec v_{\mathrm r \perp} =
\frac{1}{ 2 \pi \sigma_{\mathrm{sl}}^{2}}
e^{-\frac{(\vec v_{\mathrm{r \perp}}-\vec{A}_{\perp})^{2}}
{2 \sigma_{\mathrm{sl}}^{2}}}
v_{\mathrm{r \perp}} ~ dv_{\mathrm{r \perp}}~ d \theta}~.
\label{gg}
\end{equation}
Taking $\alpha$ to be the angle between $\vec{A}_{\perp}$ and the normal
$\vec n$ to the microlensing tube, it results
that $\varphi = \alpha+\theta$, where  $\varphi$ is the angle
between $\vec{v}_{\mathrm{r \perp}}$ and $\vec{A}_{\perp}$.

We recall that in the pixel lensing regime the effective radius of
 the microlensing tube is a function of the source star magnitude, namely
$u_{\mathrm {th}}= u_{\mathrm {th}}(M)$.  Moreover in the following
we evaluate the differential rate taking into account an
efficiency function that depends on the impact parameter, $ \epsilon =
\epsilon(u_{\mathrm {th}})$.
Therefore, we are going to replace in eq. (\ref{dnev})
$d N_{\mathrm {ev}} $ by $\int d N_{\mathrm {ev}} /d u_{\mathrm {th}} \times
\epsilon(u_{\mathrm {th}})  du_{\mathrm {th}}$, with upper limit
$u_{\mathrm T}(M)$.

Eventually, after integration on the angular variables
$\theta$ and $\alpha$,
one obtains
the expected event number rate (events sr$^{-1}$) during
the observation time $T_{\mathrm {obs}}$
\begin{equation}
\begin{array}{l}
\displaystyle{\frac{ dN_{\mathrm{ev}}}{d \Omega}} =
T_{\mathrm {obs}} ~
\displaystyle{
4 \sqrt{2} \sigma_{\mathrm {sl}}
\sqrt{ \frac { 4GM_{\odot}}{c^{2}}}
\int_0^{u_{\mathrm T}(M)} du_{\mathrm th}} ~~ \\ \\
\displaystyle{
\int_{M_{\mathrm{min}}}^{M_{\mathrm{up}}} {\phi}_{\mathrm s}(M) dM ~
\int_{\mu_{\mathrm{min}}}^{\mu_{\mathrm{up}}} d\mu ~\mu^{1/2}~
{\psi}_0(\mu)} ~ \int_{0}^{\infty} D^2_{\mathrm{os}} dD_{\mathrm{os}}
\\ \\ \nonumber
\displaystyle{
\int_{0}^{D_{\mathrm{os}}} dD_{\mathrm{ol}}
\sqrt{\frac{D_{\mathrm{ol}}(D_{\mathrm{os}}-D_{\mathrm{ol}})}
{D_{\mathrm{os}}}}
\left( \frac{     \rho_{\mathrm l}(D_{\mathrm {ol}})}
                 {\rho_{\mathrm l}(0)               } \right)
\left( \frac{ {\cal L}_{\mathrm s}(D_{\mathrm {os}})}
            { {\cal L}_{\mathrm s}(               0)}           \right)}
 ~ \\ \\ \nonumber
\displaystyle{
\int_0^{\infty}  z^2 e^{-(z^2+ \beta^2)} I_0(2\beta z) ~
\epsilon (t_{1/2},\Delta f)~dz ~.} \nonumber
\label{1234}
\end{array}
\end{equation}
where
      $z = v_{\mathrm r \perp}  /(\sqrt{2} \sigma_{\mathrm {sl}})$,
      $\beta = |\vec{A}_{\perp}|/(\sqrt{2} \sigma_{\mathrm {sl}})$
      and $I_0(2 \beta z)$ is the zero-order modified
Bessel function \footnote{By comparing eq. (\ref{1234}) with eqs.
(11) and (12) in \cite{ingrosso06} one can see that the
composition of the two maxwellian (projected) velocity
distributions for lenses and sources permits now to evaluate
analytically the two-dimensional integration on the source
velocity in eq. (12).}.

In the previous equation we explicitely take into account
an experimental event detection efficiency
$\epsilon (t_{1/2},\Delta f)$, given as a function of
the full-width half-maximum event duration $t_{1/2}$
\begin{equation}
\label{3838}
\begin{array}{l}
t_{1/2} = t_E f(a) ~,~~~~~~~~a=A_{\mathrm {max}}-1\\
              f(a) = 2 \sqrt{2}
\left(
\displaystyle{
\frac{a+2}{\sqrt{a^2+4a}}-\frac{a+1}{\sqrt{a^2+2a}}}
\right)^{1/2}~,
\nonumber
\end{array}
\end{equation}
and of the maximum flux difference during a microlensing event
\begin{equation}
\Delta f = f_0 (A_{\mathrm {max}}-1)~.
\end{equation}
Here $t_E$
is the Einstein time, $A_{\mathrm {max}}= A_{\mathrm {max}}(u_{\mathrm {th}})$
the amplification at maximum and $f_0$ the unlensed source flux.

It is well known that self-lensing and dark-lensing events
may have different time durations, depending on the MACHO mass value.
On the other hand, in pixel-lensing observations
experimental results
are usually given in terms of the $t_{1/2}$ time scale.
Thus, it is important to evaluate the expected
event rate as a function of $t_{1/2}$.

From eq. (\ref{3838}) and the relation
$t_E= R_E/ v_{\mathrm l \perp}$ it follows
\begin{equation}
t_{1/2} = \frac{ R_E f(a)} { z \sqrt{2} \sigma_{\mathrm {sl}}}~,
\label{t12z}
\end{equation}
and it is straightforward to derive the
differential event rate
\begin{equation}
\begin{array}{l}
\displaystyle{
\frac { d^2 N_{\mathrm{ev}} } {d\Omega dt_{1/2}} (t_{1/2})
        = T_{\mathrm {obs}} ~ 8 \sigma_{\mathrm {sl}}^2 ~
\int_0^{u_{\mathrm T}(M)} du_{\mathrm th}} ~~ \\ \\
\displaystyle{
\int_{M_{\mathrm{min}}}^{M_{\mathrm{up}}} {\phi}_{\mathrm s}(M) dM ~
\int_{\mu_{\mathrm{min}}}^{\mu_{\mathrm{up}}} d\mu ~
{\psi}_0(\mu)} ~ \int_{0}^{\infty} D^2_{\mathrm{os}} dD_{\mathrm{os}}
\\ \\ \nonumber
\displaystyle{
\int_{0}^{D_{\mathrm{os}}} dD_{\mathrm{ol}}
\left( \frac{\rho_{\mathrm l} (D_{\mathrm{ol}})}{ \rho_{\mathrm l} (0)} \right)
\left( \frac{{\cal{L}}_{\mathrm s}(D_{\mathrm {os}})}
  { {\cal{L}}_{\mathrm s}(0)      } \right)}
 ~\\ \\ \nonumber
\displaystyle{
  z^4 e^{-(z^2+ \beta^2)} I_0( 2 \beta z) ~ \frac{1}{f(a)}~
\epsilon (t_{1/2},\Delta f)~,} \nonumber
\label{12345}
\end{array}
\end{equation}
where $z$ is now given in terms of $t_{1/2}$ and $A_{\mathrm {max}}$
through eq. (\ref{t12z}).

The model parameters that need to be specified are
the luminosity ${\phi}_{\mathrm s}(M)$  and  mass ${\psi}_0(\mu)$
functions,
the stellar mass distributions in M31 and MW,
the mass-to-luminosity ratios for the stellar populations in M31,
the velocity dispersion $\sigma_{\mathrm s}$ and $\sigma_{\mathrm l}$
for the source and lens populations.
Further model parameters derive from the consideration
of the existence of dark matter in both M31 and MW halos.

\begin{table*}
\caption{ The M31 disk and bulge models. Relevant parameters for
WeCapp \citep{wecapp05} and our reference model. Values of mass
density, distance, mass and velocity are given in units of
$M_{\odot}$ pc$^{-3}$, kpc, $10^{10}~M_{\odot}$ and km s$^{-1}$,
respectively. For the reference model, the symbol $-$ means that
the corresponding value as in the WeCapp model is used. In the
last row we give some relevant parameter values for the models
MEGA A in \cite{mega05}.}
\medskip
\begin{tabular}{|c||cccccc||ccccccc|}
\hline
  &       &       & bulge &   &      &          &          &     &    &  disk &       &     &          \\
\hline
model     & $\rho(0)$  &$a_0$ &  $M$  &ext$_R$ & $(M/L_R)$ & $\sigma    $ &
$\rho(0)$ & $h$ & $H$&  $M$ & ext$_R$&$(M/L_R)$ & $\sigma(2h)$ \\
\hline
WeCapp      &$3.97 \times 10^4$&                    & 4.00    & 0.36 & 2.96 &  140   & 0.20 & 6.4 &  0.3 & 3.09 &  0.68 & 0.88 &  40  \\
reference   &$4.53 \times 10^4$&$2.62 \times 10^{-3}$& 3.85 &   $-$  &  $-$   &   $-$    & $-$    & $-$   & $-$   & $-$    &  $-$    & $-$    &  $-$ \\
MEGA A      &                  &                     & 4.4  &        &  3.6 &        &      &     &  1   & 5.5  &       & 2.4  &      \\
\hline
\end{tabular}
\label{tab00}
\end{table*}

\section {Models}

\subsection{Source luminosity function}

In pixel-lensing experiments only bright
and sufficiently magnified sources can give rise to detectable
microlensing events.  Monte Carlo simulations
(e.g. \cite{ingrosso06})
allow to determine the useful range of source magnitude
$M_{\mathrm {min}} \simeq -6$ and
$M_{\mathrm {max}} \simeq  3$,
and  the threshold value for the impact parameter
$u_T(M)$.

As regards the source luminosity function $\phi_{\mathrm s}(M)$,
in the lack of precise information about
the luminosity functions in M31, we adopt the
luminosity function  derived for local stars in the Galaxy
and assume that it also holds for M31, irrespectively on the position.
In particular, following \cite{MamonSoneira},
the stellar luminosity function in the magnitude range
$-6 \le M \le 16$ is given by
\begin{equation}
\phi_s(M) = H
\frac{ 10^{\beta(M-M^*)}  }
               { [ 1+10^{-(\alpha-\beta)\delta(M-M^*)}]^{1/\delta} }~,
\label{slf}
\end{equation}
where, in the $R$-band (the observational band of the MEGA collaboration)
$M^*= 1.4$, $\alpha \simeq 0.74$, $\beta = 0.045$ and $\delta= 1/3$.
The constant $H$ in eq. (\ref{slf})
is determined via the normalization condition in eq. (\ref{normphi}), namely
\begin{equation}
\int_{-6}^{16}
\phi_s (M)~L(M)~ dM~ = \rho_{\mathrm s} (0)
\left(\frac{M}{L_R}\right)^{-1}~,
\end{equation}
where $(M/L_R)$ is the mass-to-luminosity ratio
for the source star population in the $R$-band.
Note that the normalization for the source
density distribution, eq. (\ref{3030}),
implies that the event rate does
not depend on $(M/L_R)$.

\subsection{Lens mass function}

As far as the lens mass function is concerned,
for lenses belonging to the bulge and disk star populations,
the lens mass is assumed to follow a broken power law \citep{GBF}
\begin{eqnarray}
{\psi}_0(\mu) & = &
K_1~ \mu^{-0.56}~~~{\rm for}~~\mu_{\mathrm {min}} \le \mu \le 0.59
\nonumber\\
&=&
K_2~ \mu^{-2.20}~~~{\rm for}~~0.59 \le \mu \le
\mu_{\mathrm {up}}
\label{smf}
\end{eqnarray}
where the lower limit $\mu_{\mathrm {min}} = 0.1$ and the
upper limit $\mu_{\mathrm {up}}$ is $1$
for M31 bulge stars and $1.7$ for M31 and MW disk stars.
The constants $K_1$ and $K_2$ are
fixed according to the normalization condition given by eq. (\ref{nfm}).
The resulting mean mass for lenses in the bulges and disks are
$\langle m_b \rangle \sim 0.41~M_{\odot}$ and
$\langle m_d \rangle \sim 0.51~M_{\odot}$, respectively.

We also consider steeper mass function as proposed by \cite{zoccali}
and we find that our estimate of the self-lensing event number
turns out to be rather
insensitive to the mass function choice.

For the lens mass in the M31 and MW halos we
assume the $\delta$-function approximation
\begin{equation}
{\psi}_0( \mu) = \frac { \delta(\mu-\mu_{\mathrm {h}}) }
                           {            \mu_{\mathrm {h}}  }
\label{deltamuh}
\end{equation}
and take
a MACHO mass, in solar units, $\mu_{\mathrm h}= 10^{-1},~0.5,~1$.

\subsection{Mass distributions in M31 and MW}

The visible mass distributions for the M31 bulge and disk
are derived by fitting the observed brightness profiles
given by \cite{kent89} and by further
assuming mass-to-light ratios for bulge and disk stellar
populations.
Moreover, the consideration of the M31 rotation curve data
allows us to derive the distribution of the dark matter in the M31 halo.

Here, we use a coordinate system $(x,y,z)$ centered in M31, with $x$ axis
along the disk major axis. We also assume that the disk is inclined by the
angle $i=77^0$ and that the disk
azimuthal angle relative to the near minor axis
is $\phi=38.6^0$. The position angle of the bulge is $50^0$.

We neglect the MW disk since we have verified that the expected number
of events due to lenses belonging to this mass component
is only about 1\% of the total number of M31 self-lensing events.

\subsubsection{M31 bulge}
The M31 bulge model is derived from Tab. I in \cite{kent89}
containing the bulge 3-d brightness density
in the Gunn $r$-band
and the ellipticity $\epsilon(a)$
as a function of the major-axis distance
$a$ to the M31 center.

We fit the 3-d brightness profile with a
single de Vaucouleurs $a^{1/4}$ law (reference model)
\begin{equation}
j_r (a) =  j_r(0)~ 10^{- 0.4 (7.598 a^{1/4}) }~
~~~a>a_{\mathrm {min}},
\label{1devau}
\end{equation}
with central 3-d brightness density
$j_r(0) = 9.57  \times 10^{-7}$
${ L_{\odot} }~ { {\mathrm{arcsec}}^{-3}}$
(shifting to magnitudes, eq. (\ref{1devau})
may be written in the form
$m_r(a) = 15.048  + 7.598 a^{1/4}$ mag arcsec$^{-3}$).
This model accurately fits Kent data
for $a_{\mathrm {min}} \simeq 1$ arcmin, namely in the region usually
explored by pixel lensing observations.

\begin{figure}[htbp]
\vspace{7.0cm} \includegraphics{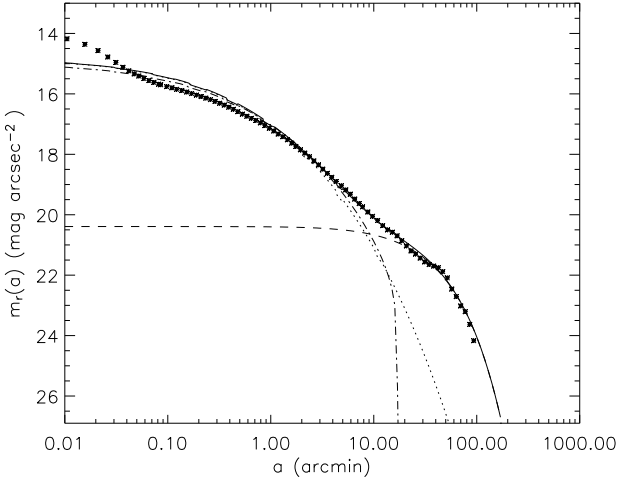}
\caption{The projected 2-d brightness profile (solid line)
is shown for the reference model in comparison with Kent data (crosses).
Dotted and dashed lines give the bulge and disk
contribution, respectively.
The dotted dashed line shows the bulge contribution for the boxy model.}
\label{fotometria2d}
\end{figure}

\begin{figure}[htbp]
\vspace{7.0cm} \includegraphics{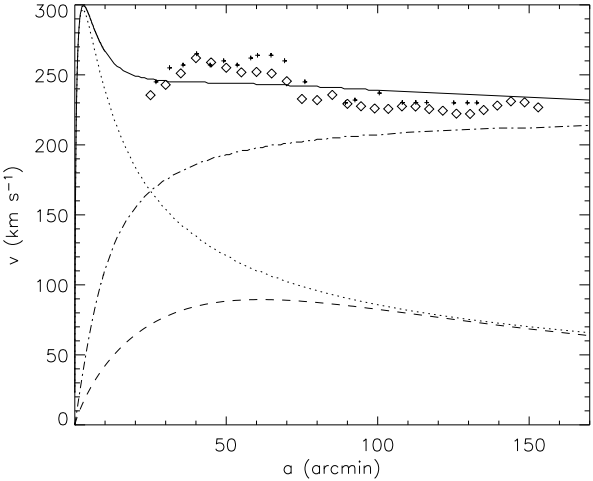}
\caption{The full M31 rotation curve (solid line) is shown in
comparison with data points derived from HI measurements of
\cite{bb84} (crosses) and \cite{carignan} (diamonds). Dotted,
dashed and dot-dashed lines give the bulge, disk and halo
contribution, respectively.} \label{curvarotnostra}
\end{figure}

From the 3-d brightness density profile in eq. (\ref{1devau}) one can derive
the corresponding mass density profile,
which has the same behaviour as the brightness profile
and central mass density
 given by
\begin{equation}
\rho(0) = \left( \frac{M}{L_R}\right)
10^{-0.4[15.048-(r-R)-{\mathrm {ext}}_R-M_{\odot R}-d_{\mathrm {mod}}]}~,
\label{3030}
\end{equation}
where
$(M/L_R)$ is the mass-to-light ratio in the $R$-band,
$(r-R)$ the color of the bulge stellar population,
$M_{\odot R}=4.42$ the absolute brightness of the Sun in the $R$-band,
${\mathrm {ext}}_R$ the extinction in the same filter and
distance modulus $d_{\mathrm{mod}} = 24.43$ (for an M31 distance
of 770 kpc).
By using the values
$(M/L_R) = 2.96$,
$(r-R)=0.59$ and ${\mathrm {ext}}_R=0.36$
quoted by \cite{wecapp05}, we obtain
$ \rho(0) = 4.53 \times 10^{4}~ M_{\odot}/{\mathrm {pc}}^3$,
corresponding to a total bulge mass
$M_{\mathrm {bulge}} \simeq  3.85 \times 10^{10}~ M_{\odot}$,
in agreement with the value given by \cite{kent89}.

Note that the observed 2-d brightness profile is also compatible with
more concentrated mass distributions for the bulge  \citep{boxymodel}.
For instance, we have tried a (boxy) model
with 99\% of the total mass inside 17.86 arcmin (4 kpc).
The mass density is now given by
\begin{eqnarray}
\rho (a) & = & 4.40 \times 10^{4} ~10^{- 0.4 (7.598 a^{0.24}) }
~~~a \le 17.86' \nonumber \\
         & = & 1.81 \times 10^{39} 10^{- 0.4 (7.598 a^{0.90}) }
~~~a > 17.86'~~~.
\label{scarpetta}
\end{eqnarray}

In Fig. \ref{fotometria2d} the  projected  2-d brightness profile
(in units of mag arcsec$^{-2}$) is shown for both models together
with Kent data.
In deriving these profiles,
we assumed that the bulge isophote are triaxial ellipsoids
with semi-major axes
\begin{equation}
a^2(\epsilon) = x^2 + y^2 + \frac {z^2} {(1-\epsilon)^2}
\end{equation}
and ellipticity varying on the semi-major axis
according to the Kent data
\footnote{
The existing relation between $\epsilon$ and $a$
may be approximated by \citep{wecapp05}
\begin{equation}
\left( \frac{1}{1-\epsilon(a)} \right)^2  =
0.254 \frac{a}{\mathrm{arcmin}}+1.11~.
\end{equation}
}.
From Fig. \ref{fotometria2d}, one can see that
beyond 0.03 arcmin
both reference and boxy
models accurately reproduce Kent data.
The only difference is the behaviour of
the bulge contribution at large distance where
in any case the disk contribution
is dominant.

For comparison we also discuss the results obtained by using the
bulge model adopted by the WeCapp collaboration \citep{wecapp05}.

\subsubsection{M31 Disk}
As in \cite{kerins01}, the disk 3-d brightness density
in the $r$-band is modeled by the law
\begin{equation}
j_r (x,y,z) = j_r (0) ~
\exp(-\sqrt{x^2+y^2}/h) ~
{\mathrm {sech}}^2(z/H)~,
\label{disk}
\end{equation}
and a best fit procedure to the Kent data (for $a \ut > 6 $ arcmin)
allows to obtain the central  brightness density
$j_r (0) = 4.2 \times 10^{-13} ~ L_{\odot} ~ { \mathrm{arcsec}}^{-3}$
(corresponding to a central magnitude $m_r(0)= 20.5$),
the radial scale length $h= 27.95$ arcmin
and the vertical scale
length $H =1.34$ arcmin (corresponding to $h=6.4$ kpc and $H=0.3$ kpc,
respectively).

As for the bulge, the corresponding disk mass density profile follows
the same behaviour as the brightness profile.
Accordingly, the disk central mass density is derived by
assuming the following parameter values
$(M/L_R) = 0.88$,
$(r-R)=0.54$ and
${\mathrm {ext}}_R=0.68$ for the disk
\citep{wecapp05}, implying
$\rho(0) = 0.2 ~ M_{\odot}~ {\mathrm {pc}}^{-3}$
and a total disk mass $M \simeq 3.09 \times 10^{10}~M_{\odot}$.
The 2-d disk brightness profile is also shown in Fig. \ref{fotometria2d}
(dashed line).

\subsubsection{M31 and MW halos}
Both M31 and MW halo mass distributions are
modeled as isothermal spheres
\begin{equation}
\rho(r) =
\displaystyle{\frac{\rho_0} {1+\left(\frac{r}{r_0}\right)^2}}~.
\label{dmm31}
\end{equation}
For M31 a fit to the M31 rotational curve
by using the three  (bulge, disk and halo) component model
allows to get the best fit parameter values
$r_0=2$ kpc and  $\rho(0)=0.23 ~M_{\odot}$ pc$^{-3}$
(see also \cite{kerins01} and \cite{wecapp05}).
The overall M31 rotational curve
and the contributions of the three components is shown
in Fig. \ref{curvarotnostra}.
In comparison with the recent determination of the mass distribution in M31
\cite{carignan}, we find that at $R=35$ kpc
the dark matter mass is $M_{\mathrm h} = 3.7 \times 10^{11}~M_{\odot}$
and the stellar mass $M_{\mathrm {vis}} = 6.6 \times 10^{10}~M_{\odot}$.
This translates in a total dynamical mass of
$\simeq 4.4 \times 10^{11}~M_{\odot}$ and
in a rotational velocity of 233 km s$^{-1}$ at $R=35$ kpc,
in agreement with the recent observations.
The M31 halo is truncated at $R=150$ kpc.

For the MW we use a core radius $a \simeq 5.6$ kpc and
a local ($R_0 = 8.5$ kpc) dark matter density
$\rho (R_0) \simeq 1.09 \times 10^7~M_{\odot}$ kpc$^{-3}$.
The corresponding asymptotic rotational velocity
is $v_{\mathrm {rot}} \simeq 220$ km s$^{-1}$. The MW halo is truncated
at $R \simeq 100 $ kpc.

\begin{figure}[htbp]
\vspace{7.0cm} \includegraphics{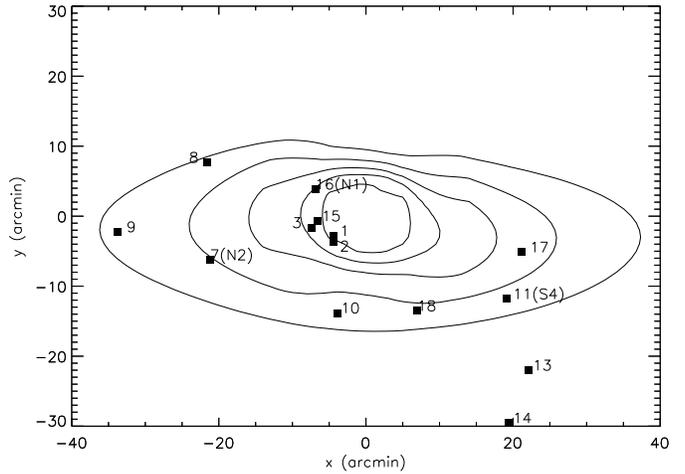}
\caption{The map $dN_{\mathrm {ev}}/d\Omega$ of the expected
(total) event rate towards M31 is shown,
assuming the reference model,
a MACHO mass value $\mu_{\mathrm h} = 0.5$
and      a MACHO halo dark matter fraction $f_{\mathrm h}=0.2$.
Here and in the following figures and tables
we adopt the observational parameters of the MEGA collaboration.
Accordingly, we consider $T_{\mathrm {obs}} = 2$ yr
and we account for the detection efficiency
$\epsilon (t_{1/2},\Delta f)$ and
maximum impact parameter $u_T(M)$ as given by \cite{mega05}.
From the  outer to the inner M31 region, contour levels
correspond to the values
             $5 \times 10^{-3}~,
             ~1 \times 10^{-2}~,
             ~2 \times 10^{-2}~,
             ~3 \times 10^{-2}~,
             ~1 \times 10^{-1}~$
event arcmin$^{-2}$, respectively.}
\label{fig1}
\end{figure}

\begin{figure}[htbp]
\vspace{7.0cm} \includegraphics{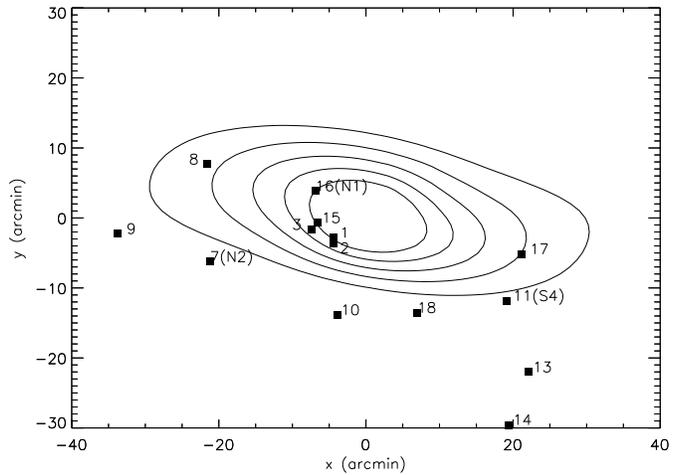}
\caption{ The same as in Fig. \ref{fig1} but for the dark-to-total
event number ratio. From the inner to outer region, contour levels
correspond to the values $0.4~,0.5~,~0.6,~0.7$ and $0.8$,
respectively. }
\label{fig2}
\end{figure}

\begin{table}[h]
\caption{The MEGA event detection efficiency $\epsilon
(t_{1/2},\Delta f)$ is given as a function of $1/\Delta f$ (first
row) for different values of $t_{1/2}$ (first column) in days. The
numerical values are derived from Fig. 12 in \cite{mega05}.}
\medskip
\begin{tabular}{||c||cccccccc||}
\hline
  &  0.02 & 0.04 & 0.08 & 0.12 & 0.16 & 0.20 & 0.24 & 0.28 \\
\hline
 1& 0.09  & 0.08&  0.02 & 0.01&     &     &     &      \\
 3& 0.22  & 0.20&  0.12 & 0.09&0.04 &0.02 &     &      \\
 5& 0.24  & 0.21&  0.14 & 0.08&0.04 &0.02 &0.01 &      \\
10& 0.29  & 0.30&  0.30 & 0.19&0.12 &0.06 &0.02 &0.01  \\
20& 0.25  & 0.25&  0.24 & 0.20&0.14 &0.09 &0.05 &0.02  \\
50& 0.14  & 0.16&  0.22 & 0.21&0.19 &0.15 &0.12 &0.08  \\
\hline
\end{tabular}
\label{tabeff}
\end{table}

\begin{table}[h]
\caption{The integrated number of expected events inside each
iso-rate contour of Fig. \ref{fig1} is given for self-lensing and
dark-lensing, assuming the reference model with $\mu_{\mathrm h} =
0.5$ and $f_{\mathrm h}=0.2$. In the last column we show the
corresponding number of events detected by the MEGA
collaboration.}
\medskip
\begin{tabular}{||c||c||c|c||c||}
\hline
Events inside  & &            &            & \\
 the 8 MEGA    & &            &            &        \\
fields         & &            &            & \\
\hline
      & iso-rate contour    & self & dark & MEGA \\
      & event arcmin$^{-2}$ &         &       &          \\
\hline
          & $1 \times 10^{-1}$ & 3.80 & 0.90  & 1  \\
          & $3 \times 10^{-2}$ & 6.49 & 2.61  & 4  \\
reference & $2 \times 10^{-2}$ & 7.45 & 3.79  & 5  \\
          & $1 \times 10^{-2}$ & 8.60 & 6.77  & 6  \\
          & $5 \times 10^{-3}$ & 9.20 & 9.68  & 12 \\
\hline
          & overall            & 9.68 & 11.76 & 14 \\
\hline
\end{tabular}
\label{tabevint}
\end{table}


\subsection {Velocity dispersions}

The random velocities of
stars and MACHOs are assumed to follow Maxwellian distributions,
with one-dimensional velocity dispersion
$\sigma = 140$ and 166 km s$^{-1}$
for the M31 bulge and MACHOs, and
$\sigma = 156$ km s$^{-1}$
for the MACHOs in the MW halo.
Moreover, following \citep{widrow},
the M31 disk stars are assumed to have one dimensional
dispersion velocity decreasing towards the outer part
from the central value $\sigma(r=0) \simeq 110$ km s$^{-1}$ to
$\sigma(r=30~{\mathrm {kpc}}) \simeq 5 $ km s$^{-1}$.
In addition, a rigid rotational velocity of 40 km s$^{-1}$
has been taken into account for the M31 bulge \citep{kerins01,An}.
For the M31 disk component the full rotational velocity
as shown in Fig. \ref{curvarotnostra} (solid line)
is also considered.

\section{Results and concluding remarks}

The main purpose of the present analysis is to compare the
predictions of our model with the observational results obtained
by the MEGA collaboration \citep{mega05}. Therefore, to evaluate
the microlensing rate we reproduce the MEGA observational set up
and we make use of the event detection efficiency $\epsilon
(t_{1/2},\Delta f)$, as a function of the time duration and
amplification at maximum and the maximum impact parameter $u_T(M)$
values as given by \cite{mega05}. In Tab. \ref{tabeff} we give
typical detection efficiency values derived from Fig. 12 in
\cite{mega05}. In order to take into account the spatial variation
of the detection efficiency we use two different evaluations of
$\epsilon$ at distances smaller and larger than 11 arcmin from the
M31 center \footnote{de Jong, private communication.}. It results
that on average $\epsilon$ is respectively smaller and larger by
about 30\% of the values quoted in Tab. \ref{tabeff}.

In the following tables and figures, we assume for both M31 and MW
halos a MACHO halo dark matter fraction $f_{\mathrm h}=0.2$ as
suggested by microlensing observations towards the Magellanic Clouds
\citep{MACHO00} and pixel-lensing observations towards M31
\citep{novati05}. However, most of our results can be easily
rescaled to other values of $f_{\mathrm h}$.
In Tab. \ref{tabtab3} we consider different values for the MACHO mass:
$\mu_{\mathrm h} = 0.1~,~0.5$ and $1$ (in solar units).
Figures 3 - 6
and Tabs. \ref{tabevint}, \ref{tabtab4} and \ref{tabtab5}
are given for $\mu_{\mathrm h} = 0.5$.

Assuming the reference model for the M31 mass distribution, the
spatial distribution of the expected events is shown in Figs.
\ref{fig1} and \ref{fig2}. Here we give maps in the sky plane of
the (total) event rate and dark-to-total event number ratio,
respectively. In Tab. \ref{tabevint} we give our estimate of the
integrated number of expected events inside each iso-rate contour
of Fig. \ref{fig1}. Here and in the following we consider events
inside the 8 fields selected by the MEGA collaboration (as
reported in Fig. 15 in \cite{mega05} the innermost M31 region is
excluded). From Fig. \ref{fig1} and Tab. \ref{tabevint} one can
see that dark-lensing gives an important contribution to
pixel-lensing beyond the second (from the inner) iso-rate contour,
namely beyond $\simeq 10$ arcmin from the M31 center.

%

The expected number of self-lensing events inside the 8 MEGA fields
is given in Tab. \ref{tabtab2}
for different source and lens populations.
Here with the symbols b,
d and h we indicate sources and/or lenses in the M31 bulge, disk and
halo, respectively. Capital symbol H is used to indicate lenses in
the MW halo. In any case, the first (second) symbol refers to the
source (lens). From Tab. \ref{tabtab2} one can see that for
all the considered models (reference, boxy and WeCapp)
the total number of self-lensing events is
roughly the same (within 15\%).

As far as the reference and boxy models are concerned, we note an
increase of bulge-bulge events to compensate a decrease of
disk-bulge ones. This is expected to be due to the different
concentration of bulge mass for the two distributions. We also
note the increase of the disk-bulge events in the WeCapp model due
to the more extended bulge mass distribution.

\begin{table}[h]
\caption{Number of self-lensing events expected given the set up of
the MEGA campaign, for the different models discussed in the text.
We consider different source and lens populations.}
\medskip
\begin{tabular}{|c||ccccc||}
\hline
Events inside &       &       &     &      &     \\
 the 8 MEGA   &       &       &     &      &     \\
 fields       &       &       &     &      &     \\
\hline
              &   bb  &     bd&   db&   dd&  self \\
\hline
reference     &  4.25 &  1.17 &  3.30& 0.96& 9.68 \\
\hline
boxy          &  5.14 &  1.10 &  2.76& 0.95& 9.95 \\
\hline
WeCapp        &  4.98 &  1.34 &  4.08& 0.96& 11.37\\
\hline
\end{tabular}
\label{tabtab2}
\end{table}
\begin{table}[h]
\caption{
For the reference model, the expected number
of dark-lensing events is given for
$\mu_{\mathrm h} = 0.1, ~ 0.5,~ 1$ and
$f_{\mathrm h}=0.2$.}
\medskip
\begin{tabular}{||c||c||ccccc||}
\hline
Events inside &&& &&& \\
 the 8 MEGA   &&& &&& \\
fields        &&& &&& \\
\hline
               & $\mu_{\mathrm h}$& bh&   bH& dh& dH& dark \\
\hline
               &0.1  &  2.55&  1.04&  8.81&  3.10&  14.49 \\
reference      &0.5  &  1.96&  0.72&  6.85&  2.23&  11.76 \\
               &1    &  1.68&  0.58&  5.80&  1.82&   9.88 \\
\hline
\end{tabular}
\label{tabtab3}
\end{table}
\begin{table}[h]
\caption{ The number of self-lensing and dark M31-lensing (due to
MACHOs in the M31 halo) events is given for our two models (here
labelled reference A and boxy A) assuming $M_b = 4.4$ and $M_d =
5.5$ (in units of $10^{10}~M_{\odot}$) and $\mu_{\mathrm h} =
0.5$, $f_{\mathrm h}=0.2$. To have the same luminosity for the M31
bulge and disk here we take $(M/L_R)_b=3.38$ and $(M/L_R)_d=1.56$.
We refer to models with $M_d = 5.5$ and $H=1$ kpc as maximal disk
models. In the last row we report some results from Tab. 5 in
\cite{mega05}, for the MEGA models in the case of high (MEGA A2)
and low (MEGA A1) extinction and for a 20\% M31 MACHO halo. }
\medskip
\begin{tabular}{|c||c|c||c|c|}
\hline
Events inside &               &             &             &  \\
the  8 MEGA   &  self         & dark        &  self       &  dark \\
fields        &               & M31         &             &  M31  \\
\hline
              &   $H=0.3$     & $H=0.3$     & $H=1$       & $H=1$  \\
              &   (kpc)       & (kpc)       & (kpc)       & (kpc)  \\
\hline
reference A   & 12.4         & 8.8        & 15.5          & 8.6  \\
\hline
boxy A        & 12.7         & 8.5        & 15.5          & 8.7  \\
\hline
MEGA A2        &     $-$       & $-$         & 12.4        &  5.7 \\
MEGA A1        &     $-$       & $-$         & 14.2        &  6.2 \\
\hline
\end{tabular}
\label{tabtab4}
\end{table}

Assuming the reference model and $f_{\mathrm h}=0.2$, in Tab.
\ref{tabtab3} we give our estimate of the expected number of
dark-lensing events for several MACHO mass value. We find that the
total number of dark-lensing and self-lensing events turns out to
be roughly the same. As regards the total (self+dark+background)
number of expected events, $\sim 23$ including $\sim 1$ event due
to SN contamination (see next), it is consistent at $2\sigma$
confidence level with the $14$ candidate MEGA events assumed to
follow a Poisson distribution.

A comparison of our results with the corresponding values reported
in Tab. 5 of \cite{mega05} for low and high internal extinction
\footnote{Note that we are considering a total extinction in the
$r$-band of 0.36 mag (0.68 mag) for the bulge (disk), irrespective
of the line of sight.} shows that there is a fairly good
agreement. Indeed, to get a more meaningful comparison for the
self-lensing contribution we normalize the values for the mass of
the luminous components to those of the MEGA models (e.g. for
their models A,
 $M_{\mathrm b} = 4.4 \times 10^{10}~M_{\odot}$ and
 $M_{\mathrm d} = 5.5 \times 10^{10}~M_{\odot}$)
and use a more broadened disk ($H=1$ kpc). In Tab. \ref{tabtab4}
we report the obtained results for our models (reference and boxy,
now labelled A) with the same bulge and disk mass as in MEGA
models A, for two values of the disk scale height $H=0.3$ kpc and
$H=1$ kpc. From Tab. \ref{tabtab4} we can see that our estimate
for the (total) number of the self-lensing events are in agreement
with the \cite{mega05} prediction only when considering more
extreme (maximal) parameters for the disk component \footnote {As
concerns our estimate in Tab. \ref{tabtab4} of dark-lensing events
due to M31 halo, we obtain a larger number of events with respect
to MEGA expectations ($\simeq 9$ events instead of $\simeq 6 $
events for $m_{\mathrm h} = 0.5$ and $f_{\mathrm h} =0.2$).
However, to describe the M31 dark matter halo we are adopting a
different density law (an isothermal profile truncated at $R=150$
kpc), which is in any case consistent with the full M31 rotation
curve.}.
\begin{table}
\caption{Distribution of the number of self-lensing events with the
distance from the M31 center for several models. In the last column,
the same quantity is given for dark-lensing assuming the reference
model, $\mu=0.5$ and $f_{\mathrm h} =0.2$.}
\medskip
\begin{tabular}{|c||ccc||c|}
\hline
Events inside    &      &       &      &       \\
the 8 MEGA       & ref. &  box. & Wec. & ref.  \\
fields           &      &       &      &       \\
\hline
 d(arcmin)
                 &self  &  self &  self& dark  \\
\hline
 2 -  5 &  3.81 & 4.55   & 3.92 &  0.93 \\
 5 - 10 &  2.80 & 3.22   & 3.43 &  1.97 \\
10 - 15 &  1.32 & 1.06   & 1.73 &  2.46 \\
15 - 20 &  0.66 & 0.27   & 0.88 &  2.20 \\
20 - 25 &  0.42 & 0.17   & 0.55 &  1.78 \\
25 - 30 &  0.22 & 0.11   & 0.28 &  1.23 \\
30 - 35 &  0.09 & 0.06   & 0.12 &  0.78 \\
35 - 40 &  0.04 & 0.03   & 0.05 &  0.33 \\
\hline
\end{tabular}
\label{tabtab5}
\end{table}

\begin{figure}[htbp]
\vspace{7.0cm} \includegraphics{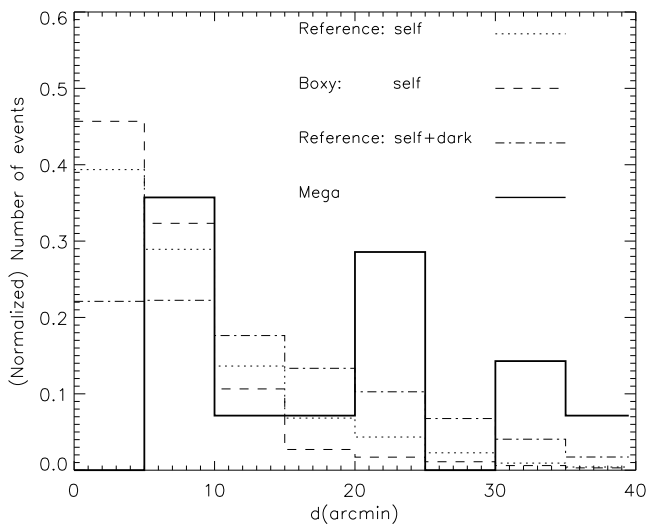}
\caption{
For the reference (dotted line) and boxy (dashed line) models,
the (normalized) distribution of the expected number of self-lensing events
within the 8 MEGA fields is given as a function of the the distance from the
M31 center.
The same quantity is shown for self+dark lensing (thin solid line)
assuming the reference model, $f_{\mathrm h}=0.2$ and $\mu_{\mathrm h}=0.5$.
For comparison the (normalized) distribution of the 14
observed MEGA events is also given (thick solid line).}
\label{fig6}
\end{figure}

Nevertheless, at variance with \cite{mega05} we do not conclude that
all the 14 events detected by the MEGA collaboration can be
explained by self-lensing only. Indeed, the spatial distribution of
the events occurring inside the 8 MEGA fields, given in Tab. \ref{tabtab5}
for several models (reference, boxy and  WeCapp) and
shown (normalized to unity)
in Fig. \ref{fig6} for both self-lensing (reference and boxy)
and total (self+dark) lensing (reference),
clearly shows that the distribution with
the distance from the M31 center of the self-lensing events hardly
can be reconciled with the MEGA data.
Indeed, as seen in Fig. \ref{fig6} an excess
of events with respect to expectations from self-lensing remains at
large distance. This conclusion is enhanced assuming the boxy
model for the M31 bulge.

\begin{figure}[htbp]
\vspace{7.0cm} \includegraphics{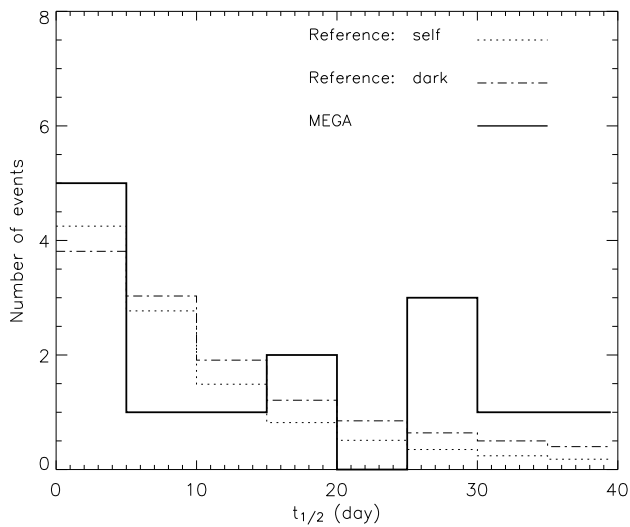}
\caption{The expected event number within the 8 MEGA fields is given
as a function of $t_{1/2}$, for both self-lensing (dotted line) and
dark-lensing (dot-dashed line) in the case of the reference model.
For comparison the distribution with $t_{1/2}$ of the 14 observed
MEGA events is also given. Here we take $f_{\mathrm h}=0.2$ and
$\mu_{\mathrm h}=0.5$.} \label{fig7}
\end{figure}

A better agreement with MEGA data can be obtained if one considers
also a dark-lensing (with $\mu_{\mathrm h} =0.5$  and $f_{\mathrm
h} =0.2$) contribution. The compatibility between the observed
MEGA event distribution as a function of distance from M31 center
and the expected one has been evaluated for both self-lensing and
self+dark lensing hypotheses\footnote{The comparison has been done
excluding one event from the MEGA candidate list (in the exterior
region) since we expect that at least one of them is due to the
contamination of background supernovae (see next for more
details.)}. By using the Kolmogorov-Smirnov test \citep{nr} we
find a K-S probability
$\simeq 0.51$ for self+dark lensing and $\simeq 0.18$ for
self-lensing only, thus implying that a dark matter contribution
to microlensing seems to be favored.


However, we caution that the candidate microlensing events could
be contamined by variable stars. In particular, the events
labelled 13 and 14, located in a region where the microlensing
rate is negligible, might be contaminated by background
supernov\ae (SN). Indeed, by assuming standard SN rate \citep{cox}
and integrating over the volume within $z_{max} \simeq 0.4$ (the
maximum distance at which the SN signal-to-noise ratio is at least
3 $\sigma$ above the typical baseline of 22 mag arcsec$^{-2}$) we
expect about one detectable SN in the outer M31 regions during the
observational MEGA campaign.

The distribution of the expected number of events with the time
scale $t_{1/2}$ is shown in Fig. \ref{fig7} for the reference model
and $\mu_{\mathrm h}=0.5$. From this figure, one can see that
self-lensing and dark-lensing events have almost the same $t_{1/2}$
distribution. Therefore, the $t_{1/2}$ event distribution is not
particularly useful to discriminate the nature of the 14 MEGA
events, at least for a MACHO mass value near $0.5~M_{\odot}$ (see
also discussion on this point in \cite{ingrosso06}).
The excess of long duration events in the MEGA data
suggests also a contamination by other variable objects.

We emphasize that our analysis shows that hardly all 14 MEGA events
can be due to self-lensing events by M31 stars.
On the other hand, given the few events detected up to now,
it seems also premature
to derive an estimate of the halo dark matter fraction in form of MACHOs.

\begin{acknowledgements}
We thank the referee for useful comments.
GI, FDP and AAN have been partially supported by MIUR through PRIN 2004
- prot. $2004020323\_ 004$.
SCN and GS have been partially supported by MIUR through PRIN 2004
- prot. $2004024710\_ 006$.
SCN and PhJ thank the Swiss National Science Foundation for support.
\end{acknowledgements}


\end{document}